\documentclass{PoS}
\usepackage[authoryear,square]{natbib}
\bibpunct{(}{)}{;}{a}{}{,}

\title{Galaxy Cluster Science with the Sunyaev-Zel'dovich Effect}

\ShortTitle{Clusters with the SZ Effect}

\author{
\speaker{Keith Grainge}$^1$, 
Stefano Borgani$^{2,3}$,
Sergio Colafrancesco$^{4}$,
Chiara Ferrari$^5$,
Anna Scaife$^6$,
Paolo Marchegiani$^4$,
S. Emritte$^4$.
J. Weller$^{7,8,9}$
\\ 
$^1$Jodrell Bank Centre for Astrophysics, School of Physics and Astronomy, The University of Manchester, Manchester, UK;\\
$^2$Astronomy Unit, Department of Physics, University of Trieste, via G.B. Tiepolo 11, I-34131 Trieste, Italy;\\
$^3$INFN-National Institute for Nuclear Physics, Via Valerio 2, I-34127 Trieste, Italy;\\
$^4$School of Physics, University of the Witwatersrand, 1 Jan Smut Ave, WITS-2050 Johannesburg, South Africa;\\
$^5$Laboratoire Lagrange, UMR 7293, Universit\'e de Nice Sophia-Antipolis, CNRS, Observatoire de la C\^ote d'Azur, 06300 Nice, France;\\
$^6$School of Physics \& Astronomy, University of Southampton, Southampton, SO17 1BJ, UK;\\
$^7$University Observatory Munich, Department of Physics, Ludwig-Maximilians University, 81679 Munich, Germany;\\
$^8$Excellence Cluster Universe, Boltzmannstr. 2, 85748 Garching bei Munchen, Germany;\\
$^9$Max-Planck-Institute for Extraterrestrial Physics, Giessenbachstrasse, 85748 Garching, Germany
\\
\\
E-mail: \email{keith.grainge@manchester.ac.uk}
}

\abstract{ Studying galaxy clusters through their
  Sunyaev-Zel'dovich~(SZ) imprint on the Cosmic Microwave Background
  has many important advantages. The total SZ signal is an accurate and
  precise tracer of the total pressure in the intra-cluster medium and
  of cluster mass, the key observable for using clusters as
  cosmological probes. Band 5 observations with SKA-MID towards
  cluster surveys from the next generation of X-ray telescopes such as
  e-ROSITA and from Euclid will provide the robust mass estimates
  required to exploit these samples. This will be especially important
  for high redshift systems, arising from the SZ's unique independence
  to redshift. In addition, galaxy clusters are very interesting
  astrophysical systems in their own right, and the SKA's excellent
  surface brightness sensitivity down to small angular scales will
  allow us to explore the detailed gas physics of the intra-cluster
  medium.}

\FullConference{
Advancing Astrophysics with the Square Kilometre Array\\
June 8-13, 2014\\
Giardini Naxos, Sicily, Italy}

\newcommand{\skipthis}[1]{}

\newcommand\apj{ApJ}
\newcommand{\apjs}{ApJS}
\newcommand{\apjl}{ApJ}
\newcommand{\aap}{A\&A}

\newcommand{\mnras}{MNRAS}

\newcommand{\araa}{ARAA}

\newcommand{\physrep}{Phys. Rept.}

\newcommand{\apss}{Astroph. Sp. Sci.}

\begin{document}

\section{Introduction}\label{intro}

\begin{figure}[h]
\includegraphics[width=0.5\columnwidth,trim={2cm 13cm 0 3cm},clip]{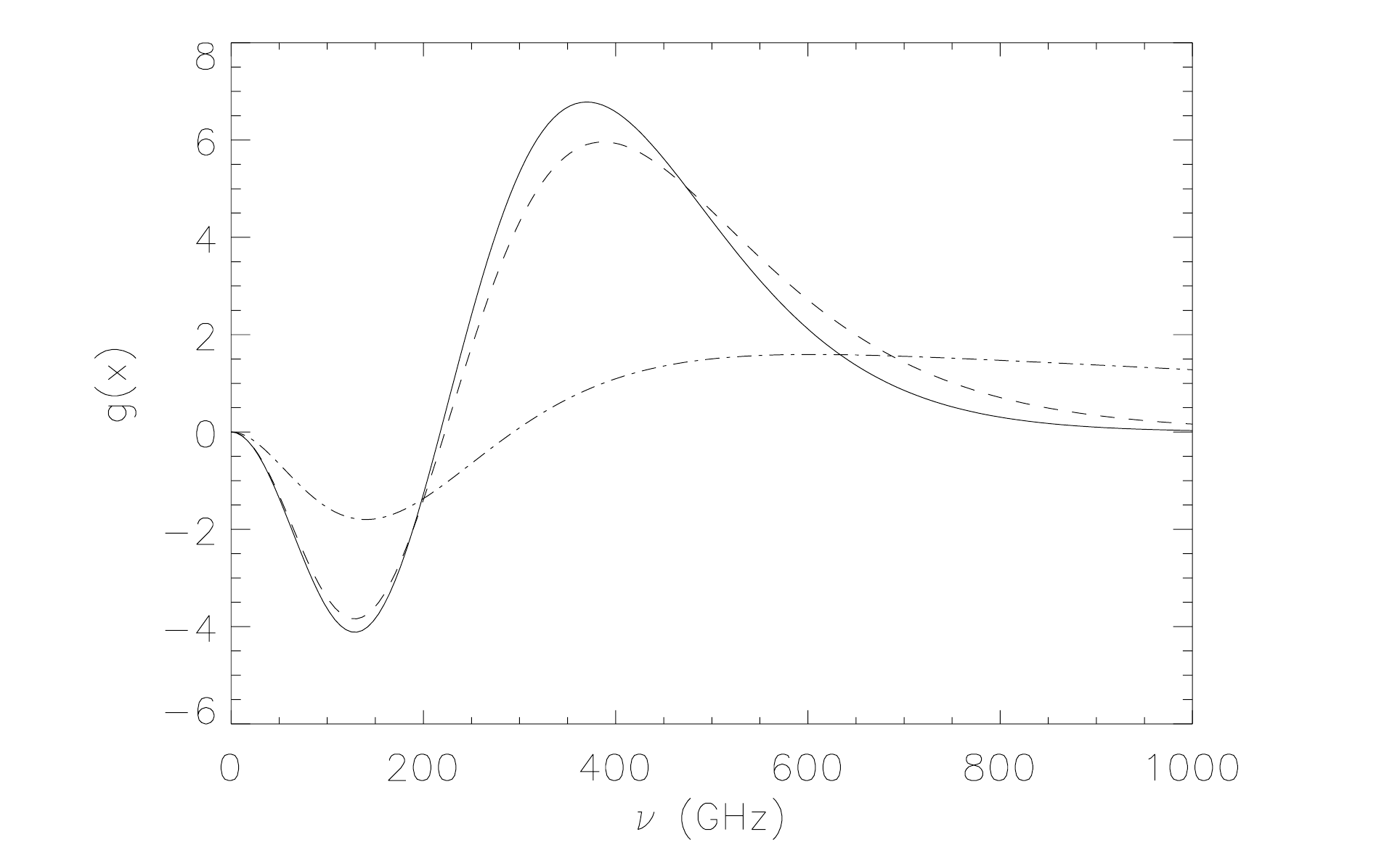}
\includegraphics[width=0.5\columnwidth,trim={2cm 13cm 0 3cm},clip]{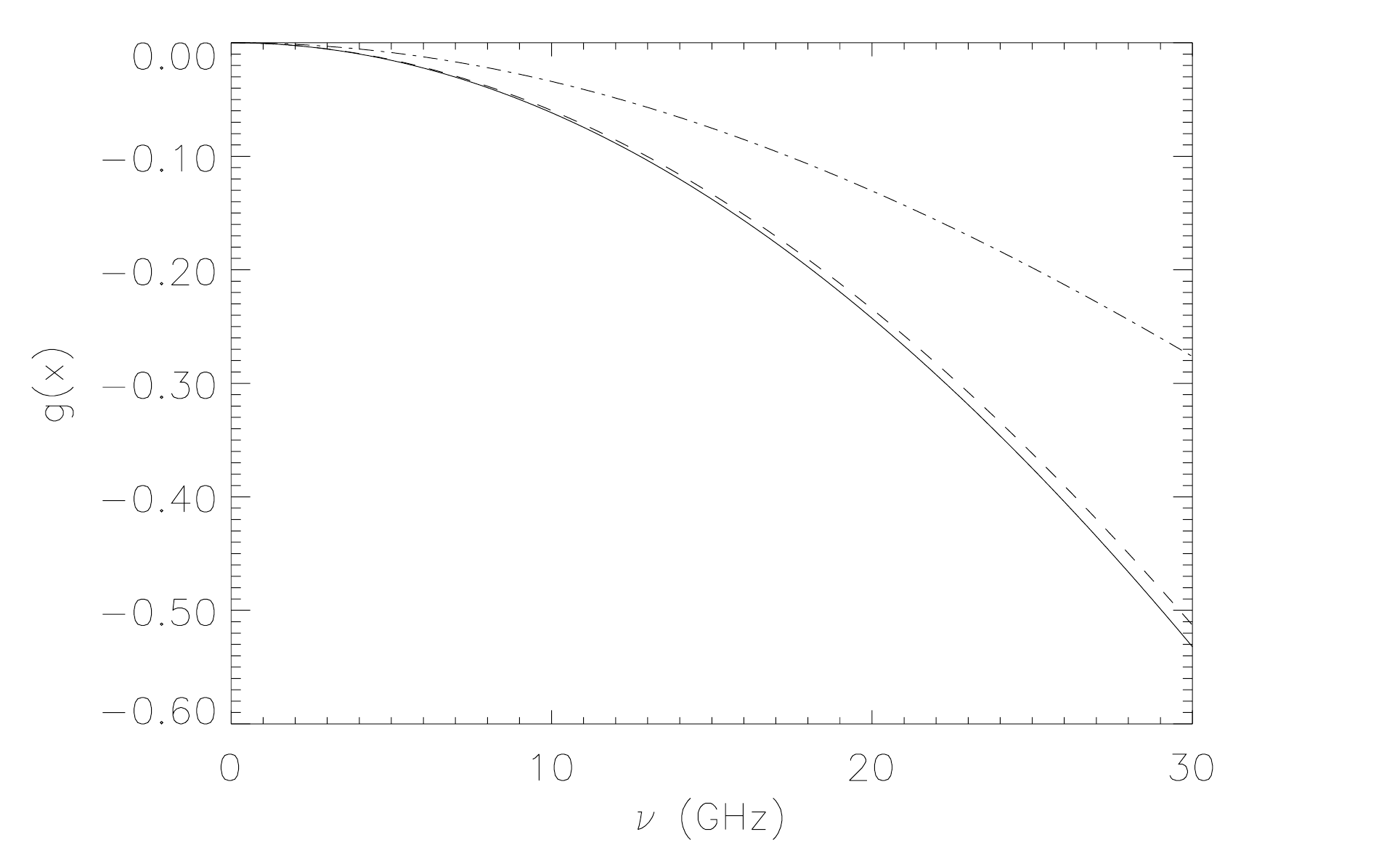}
\caption{Left. The spectral function $g(x)$ calculated for a cluster
  with temperature $k_B T_e = 10$ keV (dashed) compare to the
  non-relativistic case (solid). We also show the spectral function
  for a relativistic, non-thermal plasma with spectrum $N_{e,rel}
  \propto E^{-3.5}$ and $E_{e,min}=722.7$ keV ( dot-dashed).  Right. A
  zoom of the same plot in the frequency range up to 30 GHz, which is
  the most relevant for the SKA.~\citep{CMP2003}}
\label{fig:gx}
\end{figure}

The Sunyaev-Zel'dovich~(SZ) effect~\citep{SZ70,SZ72} is a secondary
anisotropy introduced onto the Cosmic Microwave Background~(CMB)
through inverse Compton scattering of CMB photons from the electrons
(thermal and non-thermal) contained the intra-cluster medium~(ICM) of
galaxy clusters (see e.g.~\citet{B99,CHR02} for an overview of the SZ
effect). On average this scattering leads to an increase in the energy
of these photons, while conserving photon number, which results in a
change in intensity $\Delta I_\nu$ from that of the CMB
\begin{equation}
\Delta I_\nu = {{2 (k_B T_0)^3 }\over {(h c)^2}} {\sigma _T \over
{m_e c^2}} \int P_e   g(x) dl 
\label{chaent}
\end{equation}
where $T_0$ is the temperature of the CMB today, $h$ is the Planck
constant, $k_B$ is the Boltzmann constant and $c$ is the speed of
light.  $P_e$ is the electron pressure in the cluster ICM (i.e. $P_e =
k_BT_e \cdot n_e$ for a thermal plasma with temperature $T_e$ and
number density $n_e$ of the scattering electrons); and $g(x)$ is the
spectral shape of the SZ effect (see Figure \ref{fig:gx}), in the
non-relativistic limit given by
\begin{equation}
g_{nr}(x) = {{x^4 e^x} \over { (e^x -1)^2}} \left[x \coth {x \over 2} -4 \right]
\label{gfact}
\end{equation}
in terms of the non-dimensional frequency $x$ given by
\begin{equation}
x = {{h \nu} \over {k_B T_0}} \label{xdef}
\end{equation}
SKA1-MID will have the capability to spectrally separate thermal and
non-thermal components of the SZ effect.

Equation~\ref{chaent} shows the following key aspects which make the
SZ effect a powerful probe of galaxy clusters:
\begin{itemize}
\item The surface brightness of the SZ effect is independent of
  redshift, $z$.  The cosmological dimming of radiation by $(1+z)^4$
  due to the expansion of the Universe is exactly cancelled by the
  increased CMB energy density at the time of the scattering. It is
  therefore possible to see the SZ effect from clusters all the way
  back to the epoch of their formation, provided that they retain
  their ICM.
\item The intensity of the SZ effect is proportional to the
  line-of-sight integral of the total pressure of the ICM. Therefore
  the integrated SZ flux over solid angle is proportional to the total
  thermal energy in the cluster, which is expected to be closely
  related to the overall cluster mass. This point will be returned to
  in Section~\ref{scaling}.
\end{itemize}

At frequencies below the null of the SZ effect (which is found at
217~GHz in the non-relativistic limit) the SZ effect shows an
intensity {\it decrement} towards the cluster, a very useful
characteristic signature for discriminating the SZ signal from other
emission mechanisms.  At centimetre radio frequencies we are in the
Rayleigh-Jeans regime and $I_\nu$ is well approximated as having a
thermal $\nu^2$ frequency dependence. Given that contaminating radio
halo emission typically has a falling $\nu^{-\alpha_r}$ spectrum with
$\alpha_r > 1$ \citep{feretti12}, it is often advantageous to perform SZ
work at high frequency (e.g. SPT~\citep{C11}; ACT~\citep{S11};
Planck~\citep{P11}). However, both the first reliable SZ
detection~\citep{BGH84} and the first SZ image~\citep{J93} were made at
15~GHz; and this same frequency was used by one of the next generation
of telescopes specifically designed for SZ work, AMI~\citep{Z08}, which
has made follow-up detections of 99 Planck clusters~\citep{P14}. 
Based on these observations, in this Chapter we explore the
possibility of observing the SZ-effect from galaxy clusters using SKA
operating in Band 5.  We focus upon improving knowledge of cluster
scaling relations, which are key to exploiting clusters as tracers for
cosmology; and upon probing the detailed astrophysics of the
intracluster plasma.

\section{Cluster masses with SKA SZ observations}\label{scaling}

Clusters of galaxies are recognised as powerful cosmological probes
\citep{AME11,KB12}. Measurement of their number counts constrain
cosmological parameters through the sensitive dependence of halo mass
function to the linear growth rate of cosmic density
perturbations. Furthermore, under the assumption that clusters are
fair containers of cosmic baryons, measurements of the redshift
dependence of their gas mass fraction provide tight constraints on
cosmological parameters through the expansion history.

Both such cosmological applications of galaxy clusters necessarily
require precise measurements of the total cluster mass and a precise
characterisation of the physical properties of the intra-cluster
medium. Since direct mass measurements are rather difficult to carry
out for a large ensemble of galaxy clusters, a convenient approach for
the cosmological exploitation of large cluster surveys should be based
on the measurement of suitable mass proxies which are at the same time
relatively easy to infer from observations and tightly related to
cluster mass.

As already mentioned in the Introduction, the total SZ signal of a
galaxy cluster is proportional to the total thermal energy content of
the ICM. In fact, for a cluster at redshift $z$, the integrated SZ
signal within an aperture angle $\theta$ can be written as
\begin{equation}
Y(\theta)\,=\,D_A(z)^{-2} Y(R)\,=\,2\pi D_A(z)^{-2}\int_0^R y(r)rdr\,,
\end{equation}
where $D_A(z)$ is the angular diameter distance and $y(r)$ is the
profile of the Comptonisation parameter. This quantity is proportional
to the line-of-sight integral of the electron pressure, which is
$n_eT_e$ for a thermal plasma, according to
\begin{equation}
y\,=\,\int \sigma_T n_e{k_BT_e\over m_ec^2}dl\,.
\end{equation}
Therefore, as long as the ICM can be described as a plasma in
hydrostatic equilibrium within the cluster potential well, we expect
the integrated SZ signal to be tightly related to the total cluster
mass. Under the further assumption that gas follows the DM
distribution, the prediction of the self--similar model (see, e.g.,
\citet{Arnaud10}) gives
\begin{equation}
Y_{\Delta_c}\propto E(z)^{2/3} M_{\Delta_c}^{5/3}
\end{equation} 
for the scaling relation between mass and integrated $Y$ parameter,
both computed within an aperture encompassing an overdensity of
$\Delta_c$ times the critical density at the cluster redshift,
$\rho_c(z)$. In the above equation, $E(z)$ describes the redshift
dependence of the Hubble parameter. 

\begin{figure}
\includegraphics[width=1\textwidth]{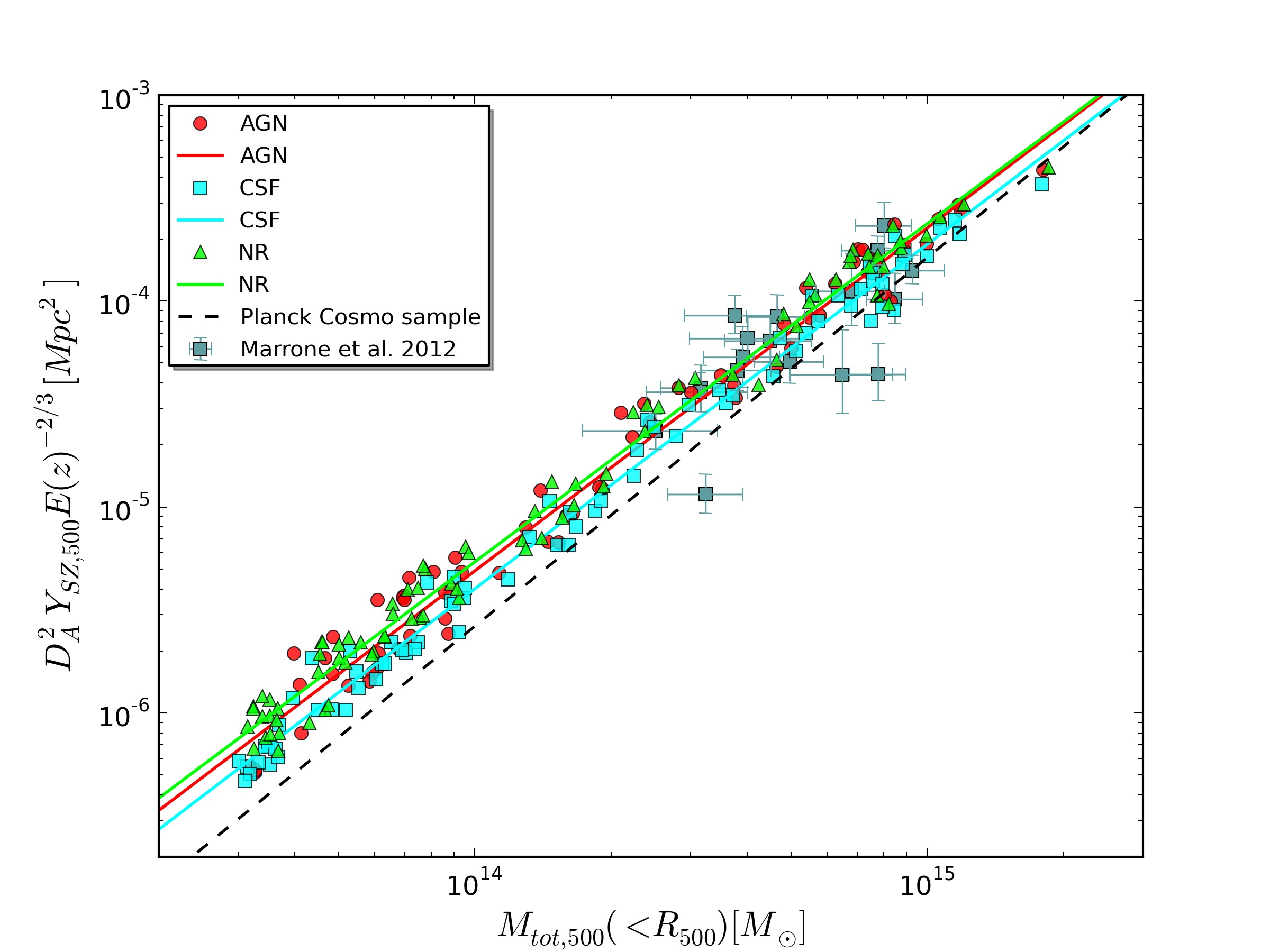}
\caption{The relation between total mass and integrated SZ signal,
  both computed within an aperture encompassing at overdensity
  $\Delta_c=500$ times the critical density, for both simulations and
  observational data (after Fabjan et al. 2014, in
  preparation). Points without errorbars refer to results from a set
  of cosmological hydrodynamical simulations of galaxy clusters
  \citep{Planelles14} (green triangles: non--radiative simulations;
cyan squares: simulations including radiative cooling, star formation
and supernova feedback; red circles: further including AGN
feedback). Gray squares with errorbars are observational results from
\citet{Marrone12}, while the dashed black line is the best-fit
 to the observational calibration presented in \citet{PlanckXX}.}
\label{fig:YM}
\end{figure}

As a consequence, the total SZ signal provides a rather
precise and robust mass proxy. It is precise, since its scaling
relation against cluster total mass is characterized by a small
intrinsic scatter. It is robust, since this scaling relation has a
quite weak dependence on the physical processes which determine the
ICM thermodynamical properties. This is illustrated in Figure
\ref{fig:YM}, which shows results on the $Y$--$M$ scaling at
$\Delta_c=500$ from a set of hydrodynamical simulations of galaxy
clusters (see also \citet{Kay12,Sembolini13,LeBrun14}), compared to
observational results (from Fabian et al. in prep). The results from
simulations clearly show that the scaling relation between integrated
SZ signal and mass has a low scatter, $<10\%$, with slope and
normalisation which are almost independent of the physical processes
included in the simulations.

Future surveys, both in the X--rays (e.g. from
eROSITA\footnote{http://www.mpe.mpg.de/eROSITA}~\citep{pillepich12})
or in optical/near-IR band (i.e.
LSST\footnote{http://www.lsst.org/lsst/} and
Euclid\footnote{http://sci.esa.int/euclid/}) will detect $\sim 10^5$
clusters and groups over an area of about $10^4$ deg$^2$. Mass
measurements in the X--rays will be limited to a small fraction (few
thousands) of mostly nearby clusters from eROSITA, while weak lensing
mass measurements in the Euclid survey will be mostly limited to $z<1$
clusters.

Thanks to its sensitivity, SKA1-MID will allow targeted follow-up
observation of clusters detected at high redshift in these future
surveys, thereby providing mass measurements, through a calibration
of the $Y$-$M$ relation, for the $z>1$ cluster population.

At the same time, the high angular resolution accessible by SKA1-MID will
allow accurate measurement of ICM pressure profiles. The verification
of the hydrostatic equilibrium condition to these pressure profiles
will allow reconstruction of mass profiles in a robust way. It is
worth pointing out that pressure profiles through X--ray observations
\citep{Arnaud10,Sun11} are obtained in an indirect way from the
combination of surface gas mass and temperature profiles. Gas clumping
is expected to bias the X-ray measurements of both gas mass profiles
from surface brightness profiles and temperature profiles because the
thermal bremsstrahlung X-ray emissivity $\varepsilon_{brem} \propto
n^2_e T^{1/2}$.  On the other hand, the tendency of gas to sit in
pressure equilibrium is such that gas clumping should have a minor
impact on pressure profiles, that are directly measured by
high-resolution SZ observations, in which the CMB temperature decrease
is $\Delta T_{CMB} \propto n_e T_e$.

\begin{figure}
\begin{center}
\includegraphics[width=8cm,angle=0]{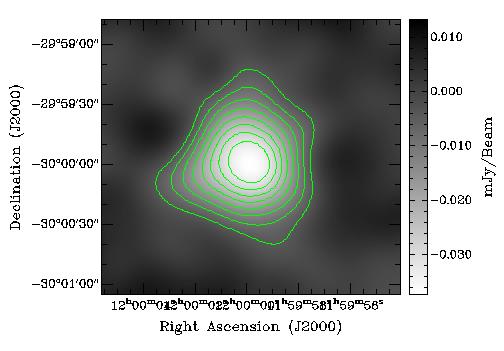}
\caption{Simulated 1-hour observation with SKA1-MID of a cluster with
  $M_{200}=4\times10^{14} \, M_O$ at $z=1.83$ mapped with a 5
  $k\lambda$ uv-taper. The cluster SZ effect is detected here at
  14~$\sigma$ c.l.
\label{z183}}
\end{center}
\end{figure}

Thanks to the different dependencies of X--ray and SZ signals on gas
density and temperature, their combination offers the possibility to
measure these two quantities without resorting to challenging X--ray
spectroscopy. This possibility is quite interesting in view of the
all--sky X-ray survey to be provided by eROSITA. Owing to the
relatively shallow flux limit of the eROSITA survey, only surface
brightness profiles will be available for most of the clusters that
will be detected in this survey at relatively high signal-to-noise
ratio. The combination of these X-ray observations with SZ maps from
SKA1-MID will allow one to recover electron pressure profiles and hence the
temperature profiles deconvolved with information coming from the
density profiles.

eROSITA is predicted to have a cluster mass detection limit of
$M_{200}=4\times10^{14} M_{\odot}$ at $z>1$ and will detect
approximately 1000 clusters at z>1 and 10 at
z>1.83~\citep{M12}. Figure~\ref{z183} demonstrates the capability of
SKA1-MID to follow-up these clusters in SZ; it shows a mock observation of
a $M_{200}=4\times10^{14} M_{\odot}$ cluster observed for one hour by
SKA-MID in band 5 (8.8--13.8~GHz) and is able to detect the SZ effect
at 14~$\sigma$. Since the SZ effect is an extended feature, we apply a
5 $k\lambda$ uv-taper as a crude matched filter. The data from the
long SKA1-MID baselines are therefore effectively discarded, but as
discussed further in Section~\ref{removal} these data are used for the
removal of contamination from radio point sources. This simualtion
demonstrates that a 1000-hour SKA1-MID programme can therefore follow up
all of the high redshift sample that eROSITA will discover. In
addition these additional observations of mass proxies will allow a
better estimate of the scatter in the mass-observable relation for
clusters detected with Euclid \citep{Rozo09}

\section{Detailed investigations of the intra-cluster medium}\label{detailed}

Deep X-ray observations of galaxy clusters have revealed different
kind of structures in the density and temperature distribution of the
ICM, from central X-ray cavities filled with the radio emitting
relativistic plasma ejected by active galaxies (e.g. \citet{fabian00}),
to high surface brightness regions, such as shock and cold fronts
related to cluster mergers (e.g. \citet{markevitch10} and references
therein). Detailed studies of the ICM pressure distribution are
necessary to characterise the complex dynamical and feedback processes
acting within galaxy clusters.

The SZ effect surface brightness provides a direct measure of the
integrated pressure along the line of sight and is well suited to
identifying ICM discontinuities (such as cold fronts or shocks) in the
absence of resolved X-ray spectroscopy. High-sensitivity and
high-resolution SZ observations are therefore an extremely valuable
tool to study the evolutionary physics of the ICM.

In the 2030 horizon, SKA2 will be ideally complemented by the Athena
X--ray satellite\footnote{http://www.the-athena-x-ray-observatory.eu}
for the study of high-redshift clusters. Thanks to its large
collecting area and spectroscopic capability, Athena will open the
possibility of studying the thermal and dynamical status of the ICM at
unprecedented precision for a significant number of galaxy clusters. A
combination of X--ray observations from Athena with detailed electron
pressure maps from SKA2 will shed light on the thermal structure of the
ICM and on the dynamics of gas motions associated to the hierarchical
build-up of galaxy clusters and by AGN feedback processes taking place
in the cluster core regions.

\begin{figure} 
\includegraphics[width=0.5\columnwidth]{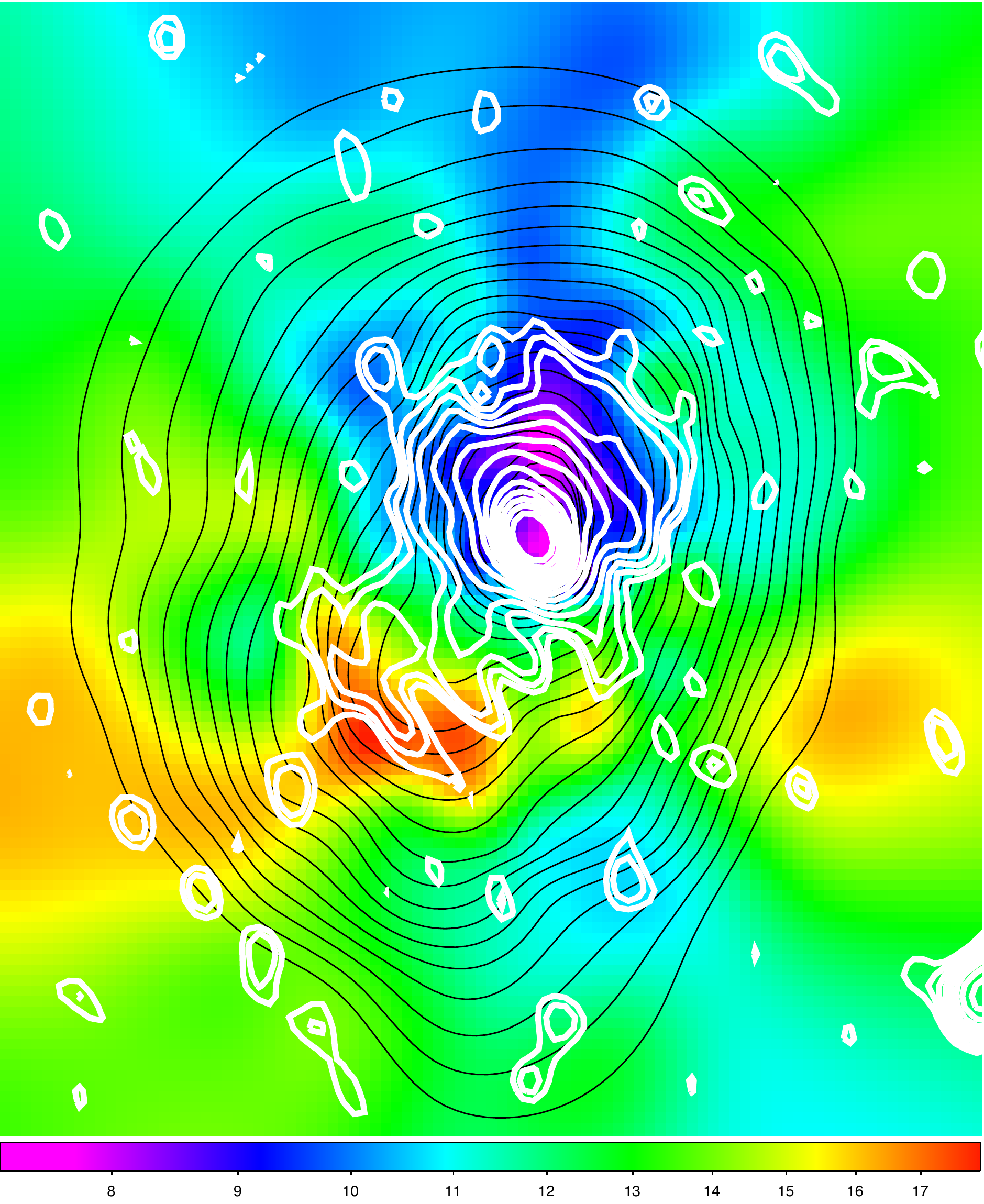}
\includegraphics[width=0.5\columnwidth]{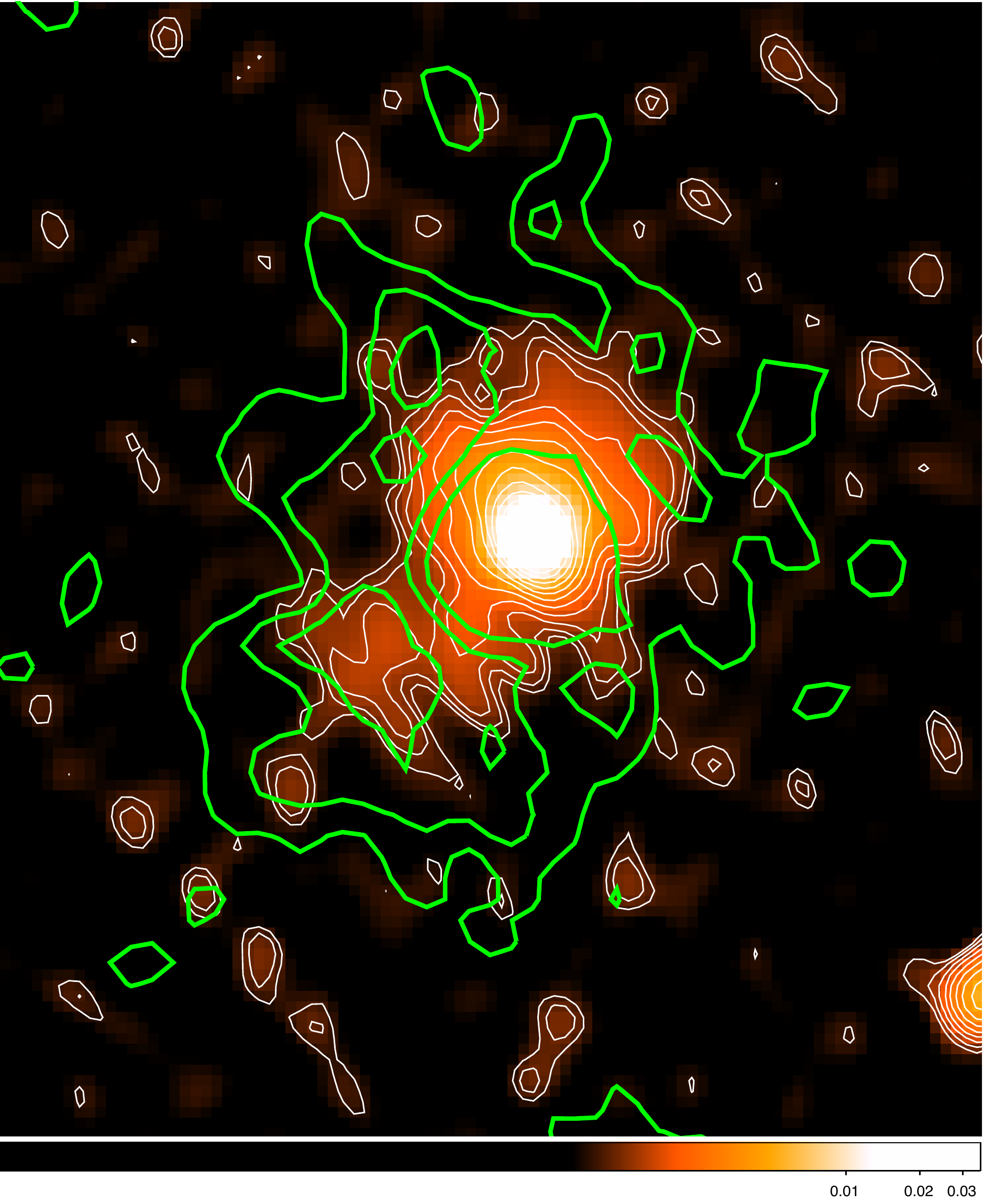}
\caption{Left: XMM-Newton X-ray temperature map of the galaxy cluster
  RXJ1347 in keV. X-ray iso-contours from the Chandra [0.5, 2.5] keV
  band image are superimposed in black. Total intensity radio contours
  are overlaid in white. They start at 3 $\sigma$ level and are spaced
  by a factor of $\sqrt{2}$. Right: total intensity 614 MHz map and
  contours (white) of RXJ1347. Contours of the MUSTANG SZE image of
  the cluster are overlaid in green (levels as in Fig.\,6 in
  \citet{mason10}). The shock region correspond to the inner contour on
  the SZ map and to the hottest (red) structure in the X-ray
  temperature map. Extracted from \citet{ferrari11}.  }
\label{fig:radio-sz-X}
\end{figure}

\begin{figure}
\includegraphics[width=6cm,angle=-90]{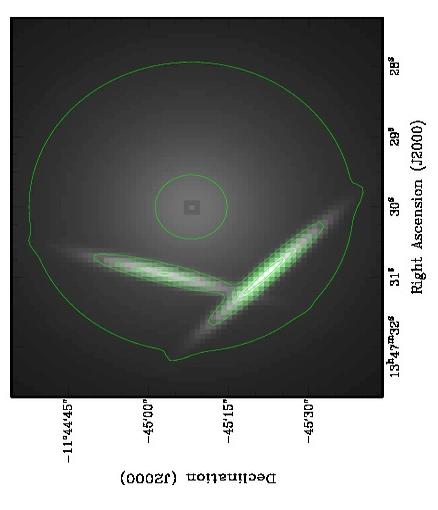}
\includegraphics[width=6cm,angle=-90]{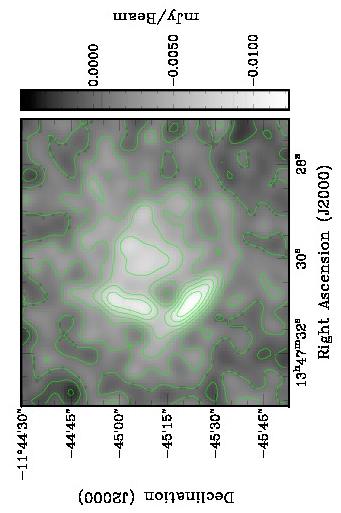}
\caption{Left: Model of cluster RXJ1347 (z = 0.451) based on
  observations at 90~GHz with the MUSTANG camera on
  GBT~\citep{mason10}. Right: Simulated 1 hour SKA1-MID observation with
  a 20~$k\lambda$ uv-taper; the southern shock heated gas region is
  detected at 20~$\sigma$. The detection of the bulk SZ effect from
  the cluster is at a significance of over 100 $\sigma$ if mapped with
  a 5~$k\lambda$ taper.\label{RXJC1347}}
\end{figure}

In the last years increasing attention is being paid to the analysis
of diffuse intracluster radio emission related to the presence of a
non-thermal component in the ICM, with the main aim of studying its
link with the complex evolutionary physics of galaxy clusters
(e.g. \citet{ferrari08,feretti12,brunetti14}). A joint analysis of
synchrotron emission and SZ signal is emerging as a promising tool to
study the interplay between the non-thermal and thermal component of
the ICM (see, e.g., \citet{basu12, cassano13, Colafrancescoetal2014}
and references therein). In this context, combined GMRT observations
(at 610 and 240 MHz) with high-resolution MUSTANG results on the
galaxy cluster RXJ1347, \citep{ferrari11} have pointed out a strong
correlation between an excess in the radio surface brightness of the
diffuse radio source at the centre of the cluster (a radio mini-halo
already detected by \citet{gitti07}) and a high pressure region
detected in the SZ map of RX J1347 and confirmed by X-ray observations
(see Fig.\,\ref{fig:radio-sz-X}). This result indicates that, in
addition to the relativistic electrons ejected by the AGN and
(possibly) re-accelerated by MHD turbulence in the central cluster
region, the presence of cosmic rays in the excess emission of the
radio mini-halo is most likely related to a shock front propagating
into the ICM. Figure~\ref{RXJC1347} shows a simulated 1-hour
observation of cluster RXJ1347 with the SKA1-MID.

These examples aim at illustrating that, if Band 5 will be covered by
the MID component, SKA will be the first telescope able to efficiently
detect both synchrotron and SZ radiation, thus becoming an extremely
powerful instrument for getting a complete view of the thermal and
non-thermal physics of the ICM.

\section{SKA1-MID observations of cluster SZ signals}

Galaxy clusters typically have a size $>1$~Mpc. The
observed SZ signal will therefore be heavily resolved out on long SKA1-MID
baselines. However, the compact core of SKA1-MID has a high filling
factor and this is critical for measuring low surface brightness
features such as the SZ effect.

The simulations presented in this chapter use only the SKA1-MID's
cross-correlation visibilities. If, in addition, auto-correlation data
are available, these will give the zero-spacing SZ flux and this will
aid immensely in inferring an accurate reconstruction of the ICM
properties. The expected statistical significance of the
auto-correlation data corresponding to the 1-hour observation in
Figure~\ref{z183} is $40\sigma$, while that in Figure~\ref{RXJC1347}
is $300\sigma$.

\section{Removal of contaminating signals}\label{removal}

It is expected that there will be substantial contamination of the SZ
signal from discrete radio sources in the field, but this
contamination can be removed by a technique described in
e.g. \citet{grainge93}. The flux and positions of the point sources
can be measured with high precision by mapping with the long SKA1-MID
baselines, which are much more numerous than the short baselines used
for measure of the SZ. As a result, for the 1 hour simulation shown in
Figure 3 the noise level from a map with a uv-taper chosen to give a
psf of $1''$ is four times lower than that in the map which shows the
SZ. Thus sources with flux $S \approx 3 \mu$Jy and above will be able to be
identified and subsequently subtracted from the short baselines that
measure the SZ. Extrapolating from the 10C source
counts~\citep{davies11} we expect to detect around 1 source per square
arcminute at this level in the field and approximately ten times as
many in the centre of the cluster. Source confusion is therefore not
an issue. The SKA1 aims to have dynamic range of 70dB in order to
address other science areas; since brightest cluster point sources are
typically $S\approx$ 10--100 mJy, for SZ work we will not need anything
like this performance.

In addition to the extended radio halos and relics continuum
emission~\citep{Ferrari15}, galaxy clusters host a variety of extended
radio sources, such as tailed radio galaxies whose shape is determined
by interaction with the ICM (e.g. \citet{GV09}); radio bubbles
that create holes in the ICM distribution and rise buoyantly through the
thermal gas (e.g. \citet{DFT05}); and compact radio sources related
to galaxy activity (either blazar and/or starburst like). In order to
study the diffuse SZ effect signal in clusters, it is of course
crucial to be able to separate the different kinds of radio
sources in the cluster environment, i.e. to discriminate between SZ and the
radio emission related to active galaxies or to other non-thermal
processes in the ICM.\\
Although these different radio sources have a wide variety of spectral
shapes, they are all very distinct from the approximately thermal
spectrum of the SZ effect (see Figure \ref{fig:gx}). Radio halo
synchrotron spectra are usually quite steep (spectral slopes
$\alpha_r>1$) and extended radio sources within the cluster (e.g.,
radiogalaxies with extended jets/lobes) also show quite steep spectra
in their extended regions due to the effects of electron aging and
radiative losses. Intra-cluster cavities have even steeper spectra
($\alpha_r \approx 0.7 - 2.3$, see \citet{Birzanetal2004}) being
filled with the termination region of radiogalaxy lobes that are
usually populated by old electrons.  A spatially-resolved spectral
analysis with the sensitivity and resolution offered by the SKA1-MID will
greatly help in providing the component separation needed to study the
extended SZ effect signal in galaxy clusters. In addition, such
spectral capability may also be able to distinguish any thermal and
non-thermal component of the SZ effect in clusters through careful
analysis of the Band 5 data from SKA1-MID.

Simulated SKA1-MID observations of non-thermal emission in galaxy
clusters~\citep{Ferrari15} demonstrate that the study of
extended radio emission features in clusters is feasible, thanks to
developments in deconvolution and source detection algorithms
optimised for the analysis of extended and diffuse radio sources.
With these developments, the SKA1-MID will allow multi-frequency images of
diffuse cluster radio sources to be made over large bands ranges
(specifically Band 5), enabling detailed spectral index studies of
galaxy clusters, which is essential for the component separation
analysis and the detection of the SZ effect signal.

This analysis can be extended in future to include i) a comparison
between deconvolution results obtained using the new reconstruction
algorithms based on compressed sensing and sparse representations
(e.g. the MORESANE algorithm and the multi-scale version of CLEAN
\citep{Dabbechetal}), ii) polarisation studies for targeted
observations, iii) an extended feasibility study taking into account
the full SKA1 frequency range (including also SKA1-LOW), iv) a detailed
analysis using the configuration of the full SKA array aimed at the
study of the SZ effect in galaxy clusters.

\section{SKA uniqueness and synergies for SZ observing}

As mentioned in Section~\ref{intro} there are several telescopes that
have been specifically designed for SZ work; the SKA1-MID will provide
useful complementary capabilities to these. The SPT and ACT telescopes
are primarily designed for surveying, enabled by their large fields of
view, but they are unattractive for pointed observational
programmes. Their angular resolution of $\approx 1'$ is insufficient
to image the detailed structure discussed in
Section~\ref{detailed}. SKA1-MID's resolution and sensitivity will be ideal
for follow-up imaging of the cluster catalogues produced by these
instruments. Similarly, Planck with its resolution of$\approx 5'$ is a
survey instrument for SZ. Its cluster catalogue has 1227
entries~\citep{planck29}, but its cluster selection function is
weighted more heavily towards low redshift than either SPT and ACT.

The SZA~\citep{C06} (now part of the CARMA array) and AMI are
centimetre wavelength interferometric telescopes that are ideal for SZ
follow-up observations. While they cannot match SKA1-MID's sensitivity,
their access to shorter baselines than those available to SKA1-MID are a
useful complement, especially for studies of low redshift clusters.

CCAT~\citep{CCAT} has resolution of a few arc-seconds and will be
capable of the same type of observations proposed in this chapter for
the SKA1-MID. \citet{golwala_study} calculates that a 600~hour
programme on CCAT could study 100 clusters in the mass range
$M=3.5\times10^{14} - 1\times10^{15} M_{\odot}$; for comparison, in
Section~\ref{scaling} we describe a 1000-hour SKA1-MID programme to follow
up the 1000 clusters with $M_{200}>4\times10^{14} M_{\odot}$in the
high redshift eROSITA sample. CCAT's great strength is its unique
frequency coverage, from 90~GHz to above 1~THz, which give the
potential for spectrally discriminating the thermal, kinematic and
relativistic contributions to the SZ effect. Combining a low frequency
8.8-13.8~GHz measurement from SKA1-MID with the CCAT measurements can
greatly help this discrimination~\citep{KHC04}.

ALMA~\citep{alma} is potentially very powerful for SZ work. However,
ALMA's field of view, even at Band 3, is small, less than $1'$, and so
it resolves out the bulk of the cluster SZ signal. For high resolution
imaging of the detailed ICM structure ALMA is, however, very
complementary to the SKA1-MID. As discussed in \citet{SG10}, a possible
future upgrade to ALMA through implementing Band 1 capabilities would
greatly improve ALMA's utility as an SZ instrument.

\section{Early SKA1 science and looking towards the full SKA}

Pointed SZ observations are an attractive early SKA1-MID science goal; it
has been shown that a great deal of good work could be performed with
MeerKAT~\citep{S09}. As has been discussed earlier, the SZ signal
is detected primarily on the shorter SKA1-MID baselines, with some
sensitivity required on longer baselines to remove radio
contamination. So assuming a build-out from the centre, filling out
the MeerKAT core to give good filling factor out to high radius
greatly enhances the SKA1-MID's for SZ work. Also, the autocorrelations
provided by the new SKA1-MID antennas give invaluable information about the
zero-spacing flux.

Looking to SKA2, a possible increase in maximum observing
frequency is extremely attractive for SZ measurements. Now that
autocorrelations are an accepted part of the SKA Baseline, one can
start to speculate whether the SKA2 will aim to fill the gap in
uv-space between the auto- and cross-correlations; following the
solution adopted by ALMA, this could be achieved with the addition of
a close-packed array of small ($\sim 7$m) dishes.

\acknowledgments

Chiara Ferrari acknowledges financial support by the ``{\it Agence
  Nationale de la Recherche}'' through grant ANR-09-JCJC-0001-01, the
``{\it Programme National Cosmologie et Galaxies (2014)}'', the {\it
  BQR} program of Lagrange Laboratory (2014). Sergio Colafrancesco
acknowledges support by the South African Research Chairs Initiative
of the Department of Science and Technology and National Research
Foundation and by the Square Kilometre Array (SKA). Stefano Borgani
acknowledges financial support from the PRIN-MIUR 2009AMXM79 Grant,
from the PRIN-INAF 2012 Grant "The Italian network for computational
cosmology" and from the INDARK INFN Grant. We would like to thank
Dunja Fabjan for providing Figure~\ref{fig:YM}. We thank Malak Olamaie
for cluster models.

\bibliographystyle{apj}

\begin{thebibliography}{99}
\bibpunct{(}{)}{;}{a}{,}{,}
\bibitem[{Allen et al.}, 2011]{AME11} Allen, S.~W., Evrard, A.~E., \& Mantz, A.~B.\ 2011,
  \araa, 49, 409 [{\tt arXiv:1103.4829}]

\bibitem[Arnaud et al., 2010]{Arnaud10} Arnaud, M., Pratt, G.~W., Piffaretti, R., et al.\
  2010, \aap, 517, A92 [{\tt arXiv:0910.1234}]

\bibitem[Basu, 2012]{basu12} Basu, K., 2012, MNRAS, 421, 112, [{\tt
    arXiv:1111.2856}]

\bibitem[Birkinshaw et al., 1984]{BGH84} Birkinshaw, M., Gull, S.~F., and Hardebeck, H., 1984,
  Nature, 309, 34

\bibitem[Birkinshaw, 1999]{B99} Birkinshaw, M., 1999, \physrep, 310, 97 [{\tt arXiv:astro-ph/9808050}]

\bibitem[Birzan et al., 2004]{Birzanetal2004} Birzan, L. et al. 2004, ApJ, 607, 800 [{\tt
    arXiv:astro-ph/0402348}]

\bibitem[Brown et al., 2004]{alma} Brown, R.~L., Wild, W., Cunningham
  C., 2004, Advances in Space Research, 34, 555

\bibitem[Brunetti \& Jones, 2014]{brunetti14}Brunetti, G., \& Jones,
  T.~W.\ 2014, International Journal of Modern Physics D, 23, 30007,
  [{\tt arXiv:1401.7519}]

\bibitem[Carlstrom et al., 2002]{CHR02} Carlstrom, J.~E., Holder,
  G.~P., and Reese, E.~D., 2002, \araa, 40, 643 [{\tt
      arXiv:astro-ph/0208192}]

\bibitem[Carlstrom, 2006]{C06} Carlstrom J. E., 2006, ASPC, 356, 35

\bibitem[Carlstrom et al., 2011]{C11} Carlstrom, J.~E. et al.,
  Publications of the Astronomical Society of the Pacific, 2011, 123,
  568 [{\tt arXiv:0907.4445}]

\bibitem[Cassano et al., 2013]{cassano13} Cassano, R. et el., 2013,
  ApJ, 777, 141, [{\tt arXiv:1306.4379}]

\bibitem[Colafrancesco et al., 2003]{CMP2003} Colafrancesco, S., Marchegiani, P. and Palladino,
  E. 2003, A\&A, 397, 27 [{\tt arXiv:astro-ph/0211649)}]

\bibitem[Colafrancesco et al., 2014]{Colafrancescoetal2014} Colafrancesco, S. et al. 2014, A\&A,
  566, 42 [{\tt arXiv:1312.1846}]

\bibitem[Dabbech et al., 2014]{Dabbechetal} Dabbech, A., Ferrari, C., Mary, D., Slezak,
  E. and Smirnov, O., 2014, in preparation.

\bibitem[Davies et al., 2011]{davies11} Davies, M. L. et al., 2011, MNRAS, 415, 2708, [{\tt arXiv:1012.3659}]

\bibitem[Dunn et al., 2005]{DFT05} Dunn, R.~J.~H., Fabian, A.~C., \& Taylor, G.~B., 2005,
  \mnras, 364, 1343 [{\tt arXiv:astro-ph/0510191}]

\bibitem[Fabian et al., 2000]{fabian00} A.\,C. Fabian, J.\,S. Sanders,
  S. Ettori, G.\,B. Taylor, S.\,W. Allen, C.\,S. Crawford, K. Iwasawa,
  R.\,M. Johnstone and P.\,M. Ogle, {\em Chandra imaging of the
    complex X-ray core of the Perseus cluster}, {\it MNRAS} {\bf 318}
  (2000) L65, [{\tt arXiv:astro-ph/0007456}]

\bibitem[Ferrari et al., 2008]{ferrari08} C. Ferrari, F. Govoni, S. Schindler, A.\,M. Bykov, and Y. Rephaeli, {\em Observations of Extended Radio Emission in Clusters}, {\it SSRv} {\bf 134} (2008) 93, [{\tt arXiv:0801.0985}]

\bibitem[Ferrari et al., 2011]{ferrari11} C. Ferrari, H.\,T. Intema, E. Orr\'u, et al., {\em
  Discovery of the correspondence between intra-cluster radio emission
  and a high pressure region detected through the Sunyaev-Zel'dovich
  effect}, {\it A\&A} {\bf 534L} (2011) 12, [{\tt arXiv:1107.5945}]

\bibitem[Ferrari et al., 2015]{Ferrari15} Ferrari, C., 2015, "Cluster Radio Halos: Imaging
  Simulations", in "Advancing Astrophysics with the Square Kilometre
  Array", Eds. R. Braun et al., Proc. of the Conference held in
  Giardini Naxos, 9-13 June 2014, in press

\bibitem[Feretti et al., 2012]{feretti12} L. Feretti, G. Giovannini, F. Govoni, and M. Murgia, {\em Clusters of galaxies: observational properties of the diffuse radio emission}, {\it A\&ARv} {\bf 20} (2012) 54, [{\tt arXiv:1205.1919}]


\bibitem[Gitti et al., 2007]{gitti07} M. Gitti, C. Ferrari,
  W. Domainko, L. Feretti and S. Schindler, {\em Discovery of diffuse
    radio emission at the center of the most X-ray-luminous cluster RX
    J1347.5-1145}, {\it A\&A} {\bf 470} (2007) 25, [{\tt
      arXiv:0706.3000}]

\bibitem[Giacintucci and Venturi, 2009]{GV09} Giacintucci, S. and
  Venturi, T., 2009, A\&A, 505, 55 [{\tt arXiv:0907.2306}]

\bibitem[Golwala, 2008]{golwala_study} Golwala, S., 2008, ``CCAT
  Feasibility Study -- SZ Science''
  http://www.astro.caltech.edu/~golwala/talks/CCATFeasibilityStudySZScience.pdf

\bibitem[Grainge et al., 1993]{grainge93} Grainge, K., Jones, M.,
  Pooley, G., Saunders, R., Edge, A., 1993, MNRAS, 265, 57

\bibitem[Jones et al., 1993]{J93} Jones, M.~E. et al., 1993, Nature, 365, 320

\bibitem[Kay et al., 1999]{Kay12} Kay, S.~T., Peel, M.~W., Short,
  C.~J., et al.\ 2012, \mnras, 422, 1999 [{\tt arXiv:1112.3769}]

\bibitem[Knox et al., 2004]{KHC04} Knox, L., Holder, G.~P., Church, S.~E., 2004, \apj,
  612, 96 [{\tt arXiv:astro-ph/0309643}]

\bibitem[Le Brun et al., 2014]{LeBrun14} Le Brun, A.~M.~C., McCarthy,
  I.~G., Schaye, J., \& Ponman, T.~J.\ 2014, \mnras, 441, 1270 [{\tt
      arXiv:1312.5462}]

\bibitem[Kravtsov \& Borgani, 2012]{KB12} Kravtsov, A.~V., \& Borgani,
  S.\ 2012, \araa, 50, 353 [{\tt arXiv:1205.5556}]

\bibitem[Markevitch, 2008]{markevitch10} M. Markevitch, {\em
  Intergalactic shock fronts}, {\it SSRv} {\bf 134} (2008) 93, [{\tt
    arXiv:1010.3660}]

\bibitem[Marrone et al., 2012]{Marrone12} Marrone, D.~P., Smith,
  G.~P., Okabe, N., et al.\ 2012, \apj, 754, 119 [{\tt
      arXiv:1107.5115}]

\bibitem[Mason et al., 2010]{mason10} B.\,S. Mason, S.\,R. Dicker,
  P.\,M. Korngut, et al., {\em Implications of a High Angular
    Resolution Image of the Sunyaev-Zel'Dovich Effect in
    RXJ1347-1145}, {\it ApJ} {\bf 716} (2010) 739

\bibitem[Merloni et al., 2012]{M12}
Merloni, A. et al., 2012, ``eROSITA Science Book''

\bibitem[Nagai et al., 2007]{NKV07} Nagai, D., Kravtsov, A. V.,
  Vikhlinin, A., 2007, {\it ApJ}, 668, 1 [{\tt
      arXiv:astro-ph/0703661}]

\bibitem[Perrott et al., 2014]{P14} Perrott, Y.~C. et al., 2014, [{\tt
    arXiv1405.5013}]

\bibitem[Pillepich et al., 2012]{pillepich12} Pillepich, A., Porciani,
  C., \& Reiprich, T.~H.\, 2012, MNRAS, 422, 44, [{\tt arXiv:1111.6587}]

\bibitem[Planck Collaboration XI, 2011]{P11} Planck Collaboration:
  Ade, P.~A.~R. et al, Astronomy \& Astrophysics, 2011, 536, 8 [{\tt
      arXiv:1101.2024}]


\bibitem[Planck Collaboration XX, 2013]{PlanckXX} Planck
  Collaboration, Ade, P.~A.~R., Aghanim, N., et al.\ 2013, [{\tt
      arXiv:1303.5080}]

\bibitem[Planck Collaboration XXIX, 2013]{planck29} Planck
  Collaboration: Ade, P.~A.~R. et al., 2013, ``Planck 2013
  results. XXIX. Planck catalogue of Sunyaev-Zeldovich sources'',
  [{\tt arXiv:1303.5089}]

\bibitem[Planelles et al., 2014]{Planelles14} Planelles, S., Borgani,
  S., Fabjan, D., et al.\ 2014, \mnras, 438, 195 [{\tt
      arXiv:1311.0818}]

\bibitem[Rozo et al., 2009]{Rozo09}
Rozo, E. et al., 2009, \apj, 699, 768 [{\tt arXiv:0809.2794}]

\bibitem[Scaife et al., 2009]{S09} Scaife, A.~M.~M. et al., 2009, COSMOKatSZ science
  proposal.

\bibitem[Scaife \& Grainge, 2010]{SG10} Scaife, A.~M.~M., \& Grainge, K.~J.~B., 2010, BASI, 38,
  185 [{\tt arXiv:1002.1895}]

\bibitem[Sembolini et al., 2013]{Sembolini13} Sembolini, F., Yepes, G., De Petris, M., et al.\
  2013, \mnras, 434, 2718

\bibitem[Sun et al., 2011]{Sun11} Sun, M., Sehgal, N., Voit, G.~M., et al.\ 2011, \apjl,
  727, L49 [{\tt arXiv:1012.0312}]

\bibitem[Sunyaev \& Zel'dovich, 1970]{SZ70}Sunyaev, R.~A. and Zel'dovich, Y.~B., 1970, \apss, 7, 3

\bibitem[Sunyaev \& Zel'dovich, 1972]{SZ72} Sunyaev, R.~A. and Zel'dovich, Y.~B., 1972, Comments on
  Astrophysics and Space Physics, 4, 173


\bibitem[Swetz et al., 2011]{S11} Swetz, D.~S. et al., 2011, \apjs,
  194, 41 [{\tt arXiv:1007.0290}]

\bibitem[Woody et al., 2012]{CCAT} Woody, D. et al., 2012, in Ground-based and Airborne
  Telescopes IV, edited by L. M. Stepp, R. Gilmozzi, and H. J. Hall,
  vol. 8444 of Proc. SPIE, 84442M

\bibitem[Zwart et al., 2008]{Z08} Zwart, J.~T.~L. et al., 2008, MNRAS, 391, 1545 [{\tt arXiv:0807.2469}]

\end{thebibliography}

\end{document}